# ARTICLE INFORMATION

**Article title**

MATLAB-Simulated Dataset for Automatic Modulation Classification in Wireless Fading Channels

**Authors**

M.M. Sadman Shafi[*], Tasnia Siddiqua Ahona, Ashraful Islam Mridha

**Affiliations**

Department of Electrical and Electronic Engineering, Islamic University of Technology, Board Bazar, Gazipur 1704, Bangladesh

**Corresponding author's email address and Twitter handle**

E-mail: sadmanshafi@iut-dhaka.edu

**Abstract**

Accurate modulation classification is a core challenge in cognitive radio, adaptive communications, spectrum analysis and related domains especially under dynamic channels without transmitter knowledge. To address this need, the article presents a labeled synthetic dataset designed for wireless modulation classification under realistic propagation scenarios. The signals were generated in MATLAB by modulating randomly generated bitstreams using five digital modulation schemes—BPSK, QPSK, 16-QAM, 64-QAM, and 256-QAM. These signals were then transmitted through Rayleigh and Rician fading channels with standardized parameters, along with additional impairments to enhance realism and diversity. Each modulated signal contains 1000 symbols. A comprehensive set of features was extracted from the signals, encompassing statistical, time-domain, frequency-domain, spectrogram-based, spectral correlation-based and image processing-based descriptors such as BRISK, MSER, and GLCM. The dataset is organized into 10 CSV files covering two channel types (Rayleigh and Rician) across five sampling frequencies: 1 MHz, 10 MHz, 100 MHz, 500 MHz, and 1 GHz. To facilitate reproducibility and encourage further experimentation, the MATLAB scripts used for signal generation and feature extraction, are also provided. This dataset serves as a valuable benchmark for developing and evaluating machine learning models in modulation classification, signal identification, and wireless communication research.

# SPECIFICATIONS TABLE

| Subject | Electrical Engineering |
|---|---|
| **Specific subject area** | Wireless Networks, Modulation Classification, Signal Processing, Machine Learning, Deep Learning. |

| Type of data | Comma Separated Values files (.csv), MATLAB scripts(.m) |
|---|---|
| Data collection | This dataset was generated using MATLAB simulations. Wireless signal data from five modulation schemes were simulated under two different fading channels (Rayleigh and Rician). After computing traditional signal-domain features, additional features are extracted using spectrogram analysis and advanced techniques such as BRISK, MSER, and GLCM. Setting different sampling frequencies (1 MHz, 10 MHz, 100 MHz, 500 MHz and 1 GHz), a total of 10 datasets (5 under each fading channel) were created. |
| Data source location | Islamic University of Technology, Board Bazar, Gazipur 1704, Bangladesh |
| Data accessibility | Repository name: Mendeley Data<br><br>Data identification number: 10.17632/kfzyp9hnzb.1<br><br>Direct URL to data: https://data.mendeley.com/datasets/kfzyp9hnzb/1 |
| Related research article | None |

## 1. VALUE OF THE DATA

- **Modulation classification under blind scenario:** The dataset consists of signal data from five broadly used digital communication schemes (BPSK, QPSK, 16-QAM, 64-QAM, 256-QAM) with 202 extracted features. The features were extracted for two different fading channels (Rayleigh and Rician). These data will be useful to develop machine learning models for modulation classification in wireless communication systems. The extracted features of this dataset were validated in [1] using supervised machine learning models. In [2], this dataset was also leveraged to train a Domain-Adversarial Neural Network (DANN), where a Gradient Reversal Layer (GRL) was employed to align feature distributions across domains. This application demonstrates the dataset's suitability for advancing deep learning frameworks designed to enhance robustness under varying channel conditions.
- **Understanding the effect of frequency variation:** As the dataset features signal data from a wide sampling frequency range (1 MHz, 10 MHz, 100 MHz, 500 MHz, and 1 GHz), it will help researchers observe the impact of variability in bandwidth on modulation scheme and classification model performances.
- **Diversity of features:** Apart from the conventional signal domain features, extracting features using spectrogram-based and image-processing techniques (BRISK, MSER, and GLCM) has diversified the dataset which will facilitate feature selection, evaluation of traditional classifiers and deep learning methods. [3-6]

- **Mimics real-world wireless environments:** By providing signal distortions due to different types of noise and fading channels, the dataset simulates practical wireless systems and aids research in cognitive radio, LTE/5G networks and adaptive communication systems.
- **Aiding regeneration of the dataset by including MATLAB scripts:** The MATLAB scripts added in the dataset folder will guide the interested individuals to reproduce the dataset under modified conditions. They will be able to extract their desired features and use it for varied purposes. Altogether it will extend the scope for further research in this field.

## 2. BACKGROUND

Classifying the modulation scheme is a principal challenge in the field of cognitive radio, interference management, adaptive communication systems and spectrum monitoring. In practical scenarios, for dynamic and unpredictable conditions, receivers have to perform modulation classification without prior information about transmitter parameters. This dataset features an extensive compilation of digitally modulated signals across five modulation schemes (BPSK, QPSK, 16-QAM, 64-QAM, 256-QAM) captured under Rayleigh and Rician fading channels, at five distinct sampling frequencies (ranging from 1 MHz to 1 GHz), effectively mimicking real-world noise and multipath effects. The incorporation of enhanced signal processing techniques like spectrogram analysis, BRISK, MSER, and GLCM-based feature extraction makes this dataset stand out from others. It will facilitate analysts to study novel machine learning and deep learning algorithms for blind modulation categorization. [7-10] The inclusion of MATLAB scripts encourages regeneration of the dataset under modified conditions. This resource will be of great help to develop advanced techniques and approaches to meet the challenge of modulation classification under blind scenarios.

## 3. DATA DESCRIPTION

The dataset is organized within a primary directory that contains three distinct sub-folders: Rayleigh Channel, Rician Channel, and MATLAB Codes. The Rayleigh Channel and Rician Channel folders include simulation data stored in CSV format, representing modulated signal features generated under their respective wireless fading channel conditions across five different sampling frequency, ranging from 1 MHz to 1 GHz. The MATLAB Codes sub-folder includes all simulation codes in .m format, corresponding to each dataset configuration. The structure of the dataset is shown in **Fig. 1.**

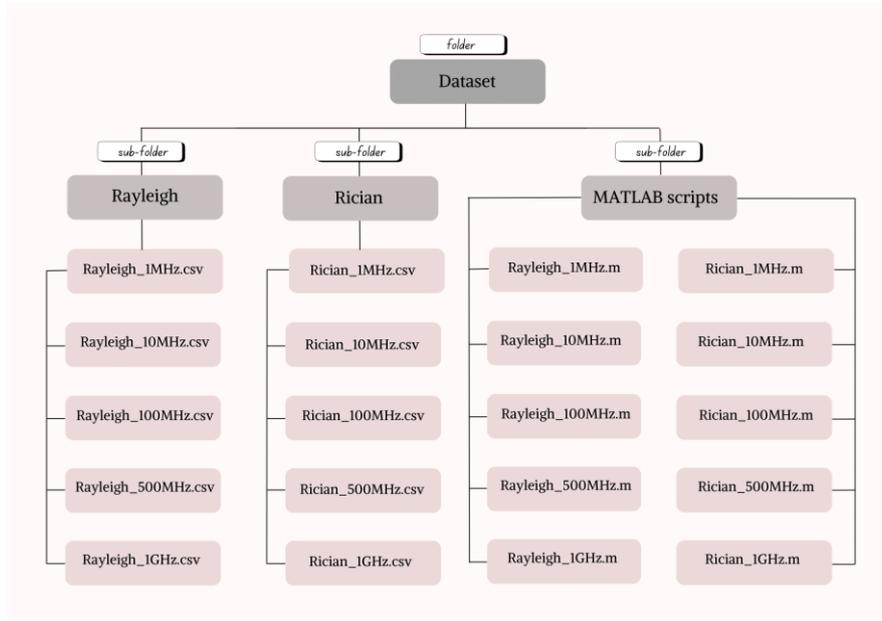

**Fig. 1.** Structure of the dataset.

Each dataset comprises 202 extracted features, systematically categorized into six groups: Statistical Moments, Time-Domain Features, Frequency-Domain Features, Spectrogram-Based Features, Spectral Correlation Features and Image Processing based Features. Among these, the first five categories—encompassing a total of 58 features—are presented in **Table 1**, which includes the exact feature names used in the dataset along with concise descriptions. These features represent various signal characteristics across time, frequency, and time–frequency domains. The remaining features are derived using image processing techniques applied to spectrogram representations, providing additional high-dimensional descriptors for signal classification and analysis.

Using image processing techniques, we extracted an additional **144 features** from the spectrogram images, including **64 BRISK features**, **64 MSER features**, and **16 GLCM features**—to effectively capture local structures, stable regions, and texture-based patterns essential for signal classification.

**Table 1**
Feature Categorization and Contextual Descriptions.

| Category | Feature Name | Description |
|---|---|---|
| Statistical Moments | MeanAmplitude | Average strength of the signal |
| | AmplitudeVariance | Fluctuation in signal amplitude |
| | AmplitudeSkewness | Asymmetry in amplitude distribution |
| | Kurtosis | Sharpness of amplitude distribution of signal |
| | SignalPower | Average energy per sample |
| | MedianAmplitude | Middle value of amplitude distribution |
| | RMSAmplitude | Effective signal magnitude |
| | StdAmplitude | Spread of amplitude values |
| | MeanDiffAmplitude | Average change between consecutive amplitudes |
| | MeanPhase | Average phase angle of the signal |
| | PhaseVariance | Spread of phase angle values |
| | EnvelopeMean | Average magnitude of the analytic signal |
| | EnvelopeStd | Spread of envelope magnitudes |
| | SpectrogramVariance | Spread of amplitude values |
| | SpectrogramSkewness | Asymmetry of amplitude distribution |
| | SpectrogramKurtosis | Sharpness of amplitude distribution of spectrogram |
| | MeanInstantaneousFreq | Central tendency of frequency variations over time |
| | InstantaneousFreqVariance | Degree of fluctuation in frequency changes |
| | InstantaneousFreqSkewness | Asymmetry in the distribution of frequency shifts |
| | InstantaneousFreqKurtosis | Sharpness or flatness of frequency distribution compared to a normal curve |
| Time-Domain Features | RSCR | Signal quality relative to noise |
| | ZeroCrossingRate | Rate of polarity changes in the signal |
| | MaxAmplitude | Highest signal magnitude |
| | MinAmplitude | Lowest signal magnitude |
| | TotalAmplitude | Sum of all amplitude values |
| | AmplitudeArea | Total accumulated amplitude |
| | AmplitudeEntropy | Uncertainty in amplitude distribution |
| | DynamicRange | Difference between maximum and minimum amplitude |
| | TotalAmpChange | Sum of all amplitude variations |
| | PAPR | Ratio of peak signal power to its average power, expressed in dB |
| Frequency-Domain Features | SignalEnergy | Total power contained in the signal |
| | SpectralCentroid | Center of mass in the frequency spectrum |
| | SpectralFlatness | Tonality measure of the spectrum |
| | SpectralPAR | Ratio of highest spectral component to mean |
| | SpectralEnergy | Total power in the frequency domain |
| | SpectralBandwidth | Width of significant spectral components |
| | LowFreqEnergy | Energy in the lower half of the spectrum |
| | HighFreqEnergy | Energy in the upper half of the spectrum |
| Spectrogram-Based Features | SpectrogramEntropy | SpectrogramEntropy |
| | SpectrogramEnergy | SpectrogramEnergy |
| | SpectrogramContrast | SpectrogramContrast |
| | SpectrogramHomogeneity | SpectrogramHomogeneity |
| | MaxSpectrogramAmplitude | MaxSpectrogramAmplitude |
| | MinSpectrogramAmplitude | Lowest intensity value in spectrogram |
| | MeanSpectrogramAmplitude | Average intensity across the spectrogram |
| | SpectralCentroidFromSpectrogram | Frequency center of energy distribution |
| | SpectralSpread | Dispersion of spectral energy around centroid |

| | LogAmplitudeSummation | Sum of log-transformed amplitude values |
| --- | --- | --- |
| | SpectralFlatnessRatio | Ratio of geometric mean to arithmetic mean of spectrum |
| | MeanMaxSpectrogram-Amplitude | Average peak intensity per time step |
| | MeanMinSpectrogram-Amplitude | Average lowest intensity per time step |
| | AmplitudeCountAboveMean | Number of spectrogram values above average |
| | MeanSpectrogramRow-Range | Average difference between max and min per row |
| | MeanSpectrogramColumn-Range | Average difference between max and min per column |
| | NormalizedSpectrogram-Energy | Average power per spectrogram element |
| Spectral Correlation Features | SCFmax | Highest spectral correlation function magnitude. |
| | AVGscfPeakFreqOffset | Mean shift in frequency where SCF reaches peak values. |
| | VARscfPeakFreqOffset | Spread of frequency shifts at SCF peak points. |

The correlation between input features and the target variable plays a critical role in the performance of machine learning classification models, as it influences both feature selection and model generalization. **Fig. 2** presents the top 30 features exhibiting the highest correlation with the target variable for one representative dataset, highlighting the most informative attributes for classification.

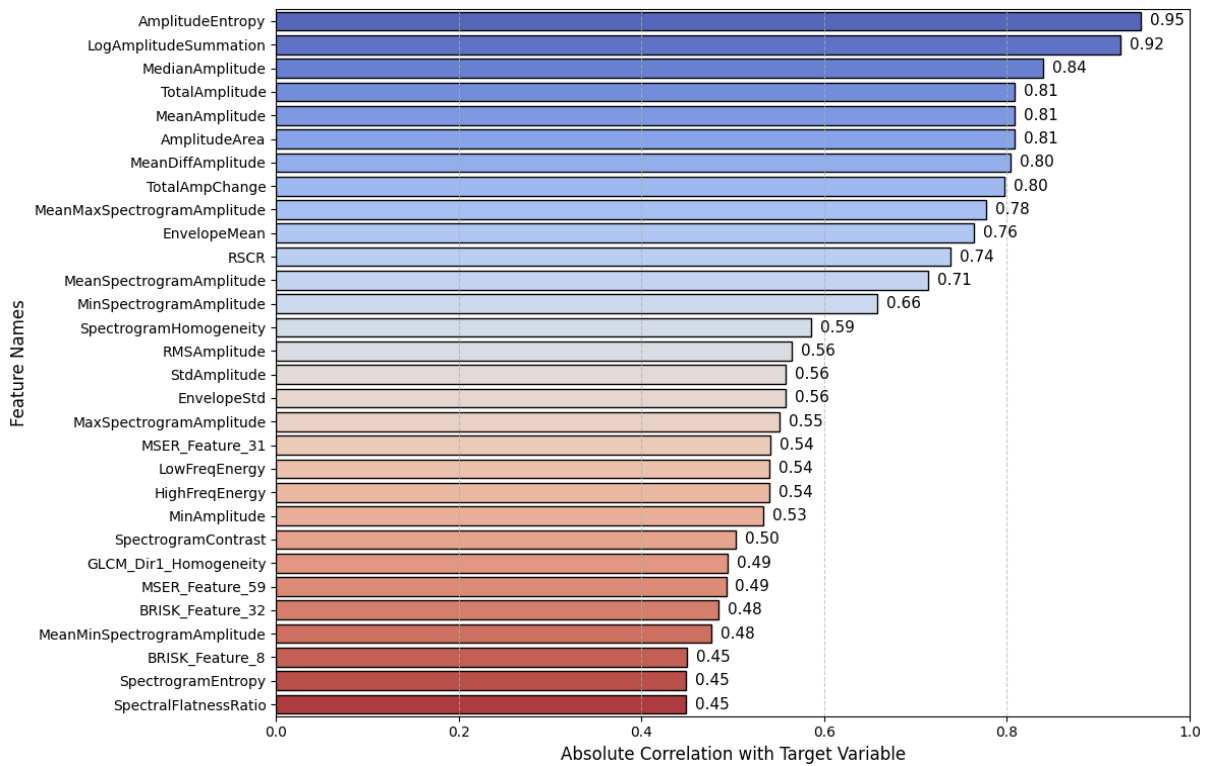

**Fig. 2.** Absolute feature correlation with target variable.

A separate folder contains the MATLAB scripts (*.m* files) corresponding to each of the 10 datasets, ensuring reproducibility and facilitating further experimentation.

In Figure 2, several high-correlation features can be explained through signal-processing perspectives. Time-domain amplitude measures such as Amplitude Entropy (uncertainty in amplitude distribution), Median Amplitude, Mean Amplitude, Total Amplitude, and Amplitude Area capture statistical variability of symbol amplitudes, which increases with higher-order QAM due to multiple constellation levels compared to constant-envelope modulations. Frequency-domain energy features such as Low-Frequency Energy and High-Frequency Energy distinguish narrower-band PSK signals from higher-order QAM schemes that spread energy across a broader spectrum. Spectrogram-based descriptors such as Log-Amplitude Summation and Spectral Flatness Ratio capture texture-like patterns in the time–frequency plane, where flatness and log-scaling highlight the difference between structured PSK spectra and more noise-like QAM distributions.

Additionally, image-derived features (BRISK, MSER, GLCM) quantify spectrogram textures: BRISK captures keypoints, MSER extracts stable intensity regions, and GLCM characterizes local contrast and homogeneity. These help separate modulation schemes by leveraging differences in spectrogram structure, such as sharp cyclic peaks for PSK versus diffuse energy patterns for QAM.

## 4. EXPERIMENTAL DESIGN, MATERIALS AND METHODS

In this portion, the detailed methodology of generating the dataset containing modulated signals with diverse features using MATLAB R2021a has been explained. The corresponding flowchart of this methodology is presented in **Fig. 3.**

**4.1 Signal Generation and Channel Simulation**

This section outlines the end-to-end process of wireless signal generation, modulation, channel modelling, and noise incorporation for building a realistic synthetic dataset suitable for classification tasks.

**4.1.1 Bitstream Generation and Modulation Schemes**

To simulate diverse wireless communication signals, random bitstreams were first generated to emulate digital transmission. These bitstreams were modulated using five commonly used modulation schemes: **BPSK, QPSK, 16-QAM, 64-QAM**, and **256-QAM**. The MATLAB functions pskmod and qammod were used for phase-shift keying (PSK) and quadrature amplitude modulation (QAM), respectively.

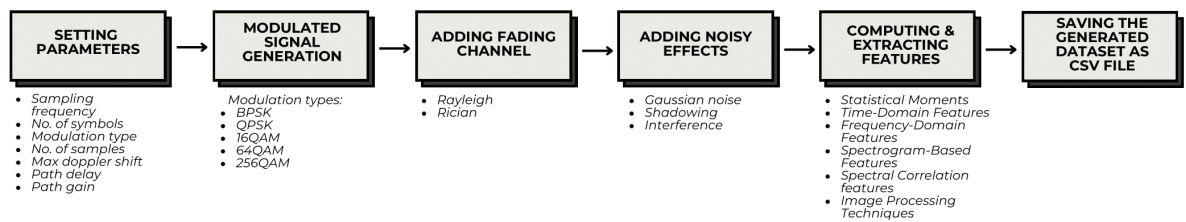

**Fig. 3.** Methodology flowchart illustrating the steps taken to construct the dataset.

- **Binary Phase Shift Keying (BPSK)**: Each bit was directly mapped to one of two distinct phases.

- **Quadrature Phase Shift Keying (QPSK)**: Bit pairs were mapped to four equidistant phase angles.
- **Quadrature Amplitude Modulation (QAM)**: Groups of 4, 6, and 8 bits were mapped to constellation points in 16-QAM, 64-QAM, and 256-QAM respectively, varying both amplitude and phase.

Each modulated signal contained 1000 symbols to maintain consistency across samples for training machine learning models. This length was selected as a practical balance between computational feasibility and statistical stability of the extracted features. 1,000 symbols were sufficient to obtain meaningful estimates for amplitude, spectral, and cyclostationary statistics across modulation orders, while keeping dataset generation time and storage manageable.

### 4.1.2 Propagation Effects via Channel Models

To reflect real-world transmission conditions, the modulated signals were passed through either a **Rayleigh** or **Rician** fading channel using MATLAB's comm.RayleighChannel and comm.RicianChannel objects.

- **Rayleigh Fading Channel:** Simulates non-line-of-sight (NLOS) scenarios common in urban and indoor environments.
- **Rician Fading Channel:** Models environments with a strong line-of-sight (LOS) component such as rural or suburban settings, with a K-factor of 6 [11]. This choice corresponds to a moderate line-of-sight (LOS) scenario, which is widely used in wireless channel modelling to represent environments with a mixture of direct and scattered components. A lower K-factor (approaching Rayleigh fading at K≈0) would increase fading severity and reduce feature stability, potentially lowering classification accuracy, while a higher K-factor would approximate AWGN conditions and generally improve classification robustness.

Both models shared the following configuration parameters:

- *Maximum Doppler Shift:* 100 Hz
- *Path Delays:* [0, 1e-6, 3e-6] seconds
- *Path Gains:* [0, -3, -6] dB

These fading effects were applied to distort the transmitted signals based on multi-path propagation and mobility.

### 4.1.3 Incorporation of Noise and Interference

To mimic the imperfections encountered in real-world wireless communication systems, various sources of noise and interference were incorporated into the signal generation pipeline. These distortions were applied cumulatively to increase the diversity and realism of the dataset.

**a) Additive White Gaussian Noise (AWGN):** To model thermal noise and background interference, AWGN was added to the signals. The noise level was dynamically adjusted using randomly selected Signal-to-Noise Ratio (SNR) values ranging from 5 dB to 30 dB, thereby simulating a wide range of channel conditions. This range is commonly used in modulation classification studies and covers moderate to good reception conditions observed in practical wireless links.

**b) Shadowing (Large-Scale Fading):** Large-scale variations in signal strength due to physical obstructions were emulated using a log-normal fading model. The signal power was attenuated following a log-normal distribution with a standard deviation of 2 dB, representing shadow fading caused by terrain or structural obstacles. Here, the value 2 dB was chosen intentionally to represent mild shadowing.

**c) Adjacent Channel Interference:** To represent interference from neighbouring frequency bands, an additional QPSK-modulated interfering signal was introduced with a frequency offset of 10 kHz relative to the main carrier. This mimics interference scenarios in congested wireless environments. While not tied to a specific standard, this offset provides a realistic interference scenario for analysis.

By combining these effects—AWGN, shadowing, and adjacent channel interference—each signal in the dataset was subjected to a realistic and challenging propagation environment, enhancing its applicability for machine learning-based classification and analysis.

**4.2 Feature Extraction**

To comprehensively characterize the modulated signals, feature extraction was performed across six categories: **Statistical Moments**, **Time-Domain Features**, **Frequency-Domain Features**, **Spectrogram-Based Features**, **Spectral Correlation Features** and **Image Processing based Features**. These features capture both structural and spectral variations in the signal, serving as the foundation for accurate classification and analysis.

**4.2.1 Statistical Moments**

Statistical descriptors were applied to various signal transformations to summarize the distribution and variability of signal characteristics. [12] The following moments were calculated:

- **Mean:** Represents the central tendency of the signal, providing an estimate of its average amplitude level over time. It serves as a baseline for analyzing variations in signal behavior.
- **Variance:** Measures the degree to which amplitude values deviate from the mean, reflecting the spread or power variability within the signal. Higher variance indicates greater fluctuation in signal strength.
- **Skewness:** Quantifies the asymmetry of the amplitude distribution. A positive skew indicates a longer right tail (more high-amplitude events), while a negative skew indicates a longer left tail (more low-amplitude events).
- **Kurtosis:** Describes the sharpness of the amplitude distribution and the presence of extreme values in the tails. High kurtosis implies the presence of sharp peaks and outliers, while low kurtosis suggests a flatter, more uniform distribution.

These moments were computed not only in the time and frequency domains but also for derived features such as the instantaneous frequency and spectrogram matrix, ensuring consistent statistical profiling across domains.

**4.2.2 Time-Domain Features**

Time-domain features were extracted directly from the raw signal amplitude to characterize its temporal behavior and modulation-specific variations. These features fall into the following categories:

- **Signal Strength and Extremes:** Features such as MaxAmplitude, MinAmplitude, TotalAmplitude, DynamicRange, TotalAmpChange, and AmplitudeArea capture both the overall magnitude and the variability of the signal. Together, they reflect the signal's peak limits, accumulated energy, and the extent of amplitude fluctuations over time—key indicators of modulation-induced behavior.
- **Complexity and Uncertainty Measures:** ZeroCrossingRate and AmplitudeEntropy quantify the rate of sign changes and randomness in the amplitude distribution, indicating the complexity and unpredictability of the waveform.
- **Power and Clarity Indicators:** PAPR (Peak-to-Average Power Ratio) evaluates power efficiency, while RSCR (Received Signal Clarity Ratio) measures signal clarity in relation to background noise—both are critical for assessing transmission quality.

### 4.2.3 Frequency-Domain Features

Extracted from the signal's Fourier spectrum, these features provide compact representations of power distribution across frequencies and aid in distinguishing modulation characteristics:

- **Energy Distribution:** Includes SignalEnergy, SpectralEnergy, LowFreqEnergy, and HighFreqEnergy, which quantify total signal power and its allocation across lower and higher frequency bands.
- **Center and Spread:** Centroid measures the center of mass of the frequency components, while bandwidth describes the range over which significant frequencies are spread.
- **Texture and Dominance:** Flatness evaluates the noisiness versus tonal quality of the frequency content, and Peak-to-Average Ratio (PAR) highlights dominant frequency components relative to the average level.

### 4.2.4 Spectrogram-Based Features

Computed from the signal's time-frequency representation, these features capture nonstationary behavior and evolving spectral patterns characteristic of the modulation schemes using the Short-Time Fourier Transform (STFT):

**STFT parameters**: window size of 128, overlap of 120 samples, and FFT size of 128.

Extracted from the spectrogram matrix:

- **Spectrogram Energy Patterns:** Features such as SpectrogramEnergy, MeanSpectrogramAmplitude, NormalizedSpectrogram-Energy, etc., quantify power distribution and intensity variation across time-frequency bins, reflecting modulation-specific energy characteristics.
- **Spectrogram Texture and Complexity:** Measures like SpectrogramEntropy, SpectrogramContrast, SpectrogramHomogeneity, etc., describe the statistical texture of the spectrogram, capturing contrast, uniformity, and structural complexity.
- **Temporal-Spectral Variation Metrics:** Features such as SpectralCentroidFrom-Spectrogram, SpectralSpread, MeanSpectrogramRow-Range, AmplitudeCountAboveMean, etc., characterize frequency dispersion, dynamic range, and local intensity changes over time.

These features emphasize localized time-frequency patterns and complement traditional signal representations. [13]

### 4.2.5 Spectral Correlation Features (Cyclostationary Analysis)

To exploit the cyclostationary nature of modulated signals, spectral correlation was analyzed:

- **Spectral Correlation Function (SCF)**: computed by circularly shifting the signal and evaluating the frequency-domain correlation between shifted versions. [14, 15]
- Derived features include:
    - **Maximum SCF value** – indicating the most prominent cyclic feature.
    - **Statistical measures of frequency offsets** – at which SCF peaks occur, reflecting modulation-induced periodicities.

These features are particularly effective in distinguishing signals with inherent periodic structures.

### 4.2.6 Image Processing Techniques

To complement statistical and spectral features, advanced image processing techniques were employed to extract robust descriptors from spectrogram representations of the modulated signals. [16, 17] These techniques—**BRISK (Binary Robust Invariant Scalable Keypoints)**, **MSER (Maximally Stable Extremal Regions)**, and **GLCM (Gray-Level Co-occurrence Matrix)**—capture distinct structural and textural characteristics inherent in the time-frequency domain.

**BRISK** (Binary Robust Invariant Scalable Keypoints) is a keypoint-based descriptor that detects salient features invariant to scale and rotation. It employs a concentric sampling pattern around each keypoint and performs pairwise intensity comparisons to generate binary strings, yielding compact and discriminative descriptors. Applied to spectrograms, BRISK captures modulation-specific local structures in the time-frequency domain, enhancing robustness to noise, geometric distortions, and signal variability [18].

To extract local structural features from the time-frequency domain, we applied the **BRISK** algorithm to the grayscale spectrogram images of the modulated signals. Using BRISK, distinctive keypoints were identified which represent unique local patterns within the spectrogram. **Fig. 4** displays the BRISK keypoints detected on the spectrogram of a generated 16-QAM signal. Following keypoint detection, BRISK descriptors were computed around each keypoint, encoding intensity variations within localized regions. To create a compact and uniform representation across samples, the mean of all BRISK descriptors per image was calculated, forming a fixed-length **briskFeatureVector**. This feature vector serves as a low-dimensional but discriminative summary of the spectrogram's content, facilitating efficient classification or clustering in downstream machine learning tasks. **Table 2** presents the key properties of the extracted BRISK features.

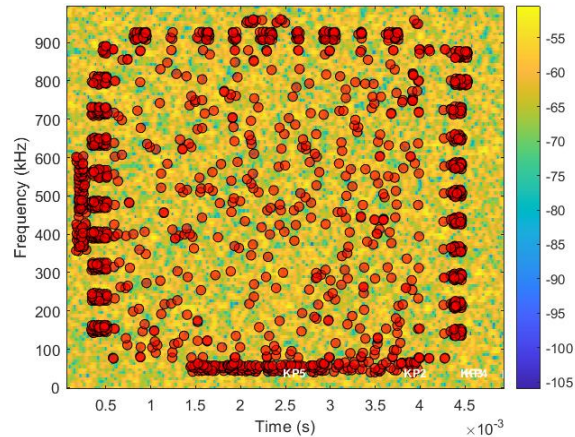

**Fig. 4.** Enhanced BRISK keypoints on spectrogram.

**Table 2**
Key properties of extracted BRISK features.

| Parameter | Size | Description |
| --- | --- | --- |
| briskPoints | 1380 keypoints | Initially detected keypoints with scale, orientation, and metric values. |
| validPoints | 1082 keypoints | Filtered keypoints used in actual feature extraction. |
| briskFeatures | 1082 × 64 matrix | Binary descriptors capturing local image patterns around valid keypoints. |
| briskFeatureVector | 64-dimensional vector | Mean descriptor representing overall BRISK features in compact form. |

**Maximally Stable Extremal Regions (MSER)** is a region-based feature detection algorithm that identifies stable, blob-like structures in grayscale images. It operates by analyzing the intensity landscape of the image and consists of the following key steps:

I. **Pixel Intensity Sorting**: Pixels are first sorted based on their grayscale intensity values using efficient non-comparison-based algorithms such as counting sort or bin sort, achieving linear time complexity.
II. **Connected Component Tracking**: As the intensity threshold is varied, connected components are formed and merged using a disjoint-set data structure and union-find algorithm to efficiently manage region connectivity.
III. **Stability Evaluation**: For each connected component, the algorithm monitors changes in region area across varying thresholds. Regions exhibiting minimal area variation over a range of intensities are deemed **maximally stable** and are selected as key features. [19]

The MSER algorithm was employed using MATLAB's detectMSERFeatures() function to extract robust features from spectrogram images. These regions, defined by intensity stability, were extracted from grayscale spectrogram images. For each detected region, geometrical and spatial properties such as pixel locations, orientation, and region axes were computed and stored. Next, the valid MSER regions were refined using SURF keypoint detection to ensure that only robust and discriminative points were retained. The combination of MSER and SURF helped filter out unstable or irrelevant regions. After identifying valid regions, a 64-dimensional descriptor was generated for each MSER zone using histogram-based feature encoding. These descriptors were then aggregated and used to construct the final **MSER feature vector** for each signal sample. This process enabled the transformation of high-resolution spectrogram images into fixed-length, compact vectors that encapsulate region-specific intensity distributions, suitable for classification or clustering. **Fig. 5** shows the MSER regions identified in a 16-QAM signal. The characteristics of MSER regions, SURF keypoints, and the feature vector are detailed in **Table 3**.

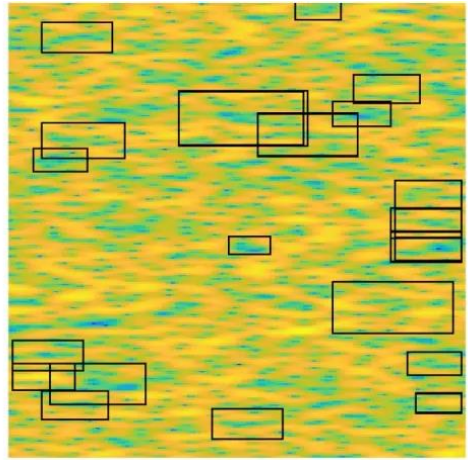

**Fig. 5.** MSER regions on spectrogram.

**Table 3**

MSER regions, SURF keypoints and feature vector properties.

| Category | Property | Value |
| --- | --- | --- |
| **MSER Regions** | Total Count | 29 |
| | Location | 29×2 single |
| | Axes | 29×2 single |
| | Orientation | 29×1 single |
| | Pixel List | 29×1 cell |
| **Valid SURF Keypoints** | Total Count | 29 |
| | Scale | 29×1 single |
| | Sign of Laplacian | 29×1 int8 |
| | Orientation | 29×1 single |
| | Location | 29×2 single |
| | Metric | 29×1 single |
| **MSER Feature Vector** | Feature Length | 29×64 single matrix |

**GLCM** is a texture analysis method that captures the spatial relationship of pixel intensities by computing co-occurrence matrices at defined offsets and directions. From these, second-order statistical features—such as contrast, correlation, energy, and homogeneity—are extracted. To improve discriminative power, features were calculated across multiple angles (0°, 45°, 90°, and 135°), capturing modulation-induced texture variations in the spectrogram.

In our approach, GLCM was applied to the grayscale spectrogram images to extract contrast, correlation, energy, and homogeneity. These features were computed using MATLAB's graycomatrix() and graycoprops() functions by analyzing multiple directional offsets to capture diverse spatial dependencies in the spectrogram. Initially, four primary directional offsets ([0,1], [-1,1], [-1,0], [-1,-1]) were selected to compute co-occurrence matrices, ensuring rotational and directional texture coverage. The feature values across these offsets, extracted from a generated 16 QAM are outlined in **Table 4**.

Each matrix captured pixel-pair statistics across a specific offset, and the derived features were averaged to form a robust descriptor set. This allowed us to quantify modulation-induced texture variations effectively. The contrast feature reflected local intensity changes; correlation measured pixel dependency; energy indicated spectral regularity; and homogeneity assessed local smoothness. These GLCM-derived descriptors were then appended to the overall feature vector, contributing to the classification and analysis of different modulation schemes with enhanced robustness and spatial sensitivity.

**Table 4**

GLCM Feature Values for Spectrogram Analysis.

| Offset | Contrast | Correlation | Energy | Homogeneity |
|---|---|---|---|---|
| (0,1) | 0.27774 | 0.8161 | 0.25053 | 0.88385 |
| (-1,1) | 0.7034 | 0.53423 | 0.16688 | 0.76937 |
| (-1,0) | 0.65162 | 0.56861 | 0.17365 | 0.78162 |
| (-1,-1) | 0.70448 | 0.53352 | 0.16616 | 0.76978 |

BRISK, MSER, and GLCM descriptors were processed into fixed-dimensional vectors (64, 64, and 16 features respectively) to ensure consistent dimensionality across all samples, which is necessary for machine learning classification. While this fixed representation compresses spectrogram textures into summary statistics, it preserves the dominant keypoint, region, and co-occurrence characteristics that remain informative under different sampling frequencies and channel conditions.

Together, the integration of BRISK, MSER, and GLCM enables a comprehensive image-based representation of the signals. These features capture both local and global structures within spectrograms, significantly enriching the feature space for downstream classification tasks. [20-22]

Since the dataset spans sampling frequencies (Fs) from 1 MHz to 1 GHz, we note that different features respond differently to changes in Fs. Amplitude-based statistics (e.g., mean amplitude, variance, skewness, RMS) are largely invariant to sampling frequency, as they depend on the symbol sequence. In contrast, frequency-domain features (e.g., spectral centroid, bandwidth, spectral spread) scale proportionally with Fs, as their frequency axis expands with higher sampling rates. Cyclostationary features (SCF offsets) shift quantitatively with Fs. Spectrogram-based features are also influenced by Fs, because the STFT window length was fixed in samples; therefore, at higher Fs, time resolution increases while frequency resolution decreases, altering spectrogram textures. Consequently, image-based features (BRISK, MSER, GLCM) derived from spectrograms also vary indirectly with sampling frequency. These variations are intentional, as they provide a testbed for evaluating how robust modulation classifiers are to differences in sampling frequency across practical systems.

# LIMITATIONS

Despite its diversity, the dataset has some inherent limitations. Limiting the dataset to five digital modulation schemes may reduce its applicability to broader communication scenarios involving diverse modulation types. The signals are synthetically generated, and while channel impairments (Rayleigh/Rician fading, AWGN, shadowing, adjacent channel interference) are modelled, hardware-specific artifacts (e.g., IQ imbalance, phase noise) and real-world spectral congestion are absent. The dataset size may be suboptimal for some deep learning architectures demanding large-scale data. This size is sufficient for classical ML methods (e.g., Random Forest, SVM, XGBoost) and lightweight deep learning models such as MLPs, shallow CNNs, or LSTMs. However, very deep architectures (e.g., ResNet, VGG, Transformers) usually require hundreds of thousands to millions of samples and may overfit on this dataset. To mitigate this, users can apply augmentation, generate synthetic expansions using GANs or signal-mixing, or use transfer learning from larger AMC datasets with fine-tuning. Additionally, the use of fixed statistical models for noise and channel dynamics may not capture temporal or spatial variations in real environments. However, the included MATLAB scripts enhance reproducibility and flexibility, allowing users to customize modulation types, sampling rates, SNR ranges, and feature sets, partially offsetting the static nature of the dataset.

# ETHICS STATEMENT
The authors confirm that they have adhered to the ethical requirements for publication in Data in Brief and the current work does not involve human subjects, animal experiments, or any data collected from social media platforms.

# CRediT AUTHOR STATEMENT
**M.M. Sadman Shafi:** Conceptualization, Methodology, Software, Writing - Review and Editing, Project Administration.

**Tasnia Siddiqua Ahona:** Validation, Formal Analysis, Data Curation, Writing – Original Draft, Visualization.

**Ashraful Islam Mridha:** Resources, Supervision, Investigation.

# ACKNOWLEDGEMENTS
This research did not receive any specific grant from funding agencies in the public, commercial, or not-for-profit sectors.

# DECLARATION OF COMPETING INTERESTS
The authors declare that they have no known competing financial interests or personal relationships that could have appeared to influence the work reported in this paper.

# DECLARATION OF GENERATIVE AI AND AI-ASSISTED TECHNOLOGIES IN THE WRITING PROCESS
During the preparation of this work the authors used ChatGPT to assist in improving the readability and language of the manuscript. After using this tool, the authors carefully reviewed and edited the content as needed and take full responsibility for the content of the published article.